# Spontaneous evolution of self-assembled phases from anisotropic colloidal dispersions

Ravi Kumar Pujala[1], Nidhi Joshi[1] and H. B. Bohidar[1,2]

[1]School of Physical Sciences, Jawaharlal Nehru University
[2]Special Center for Nanosciences, Jawaharlal Nehru University
New Delhi-110067, India

(Dated: June 7, 2013)

We investigate the spontaneous evolution of various self-assembled phase states from a homogeneous aqueous dispersion of high-aspect ratio Montmorillonite (*Na* Cloisite) nanoclay platelets grounded on the observations made over a period of 3.5 years. We have established the $t_w$-$c$ phase diagram for this system for the first time in salt-free suspensions under normal pH conditions using rheology experiments and have detected that these suspensions do undergo nontrivial phase evolution and aging dynamics. Distinctive *phase separation*, *equilibrium fluid* and *equilibrium gels* in the $t_w$-$c$ phase space are discovered. Cole-Cole plots derived from rheology measurements suggested the presence of inter connected network-like structures for $c > c_g$, $c_g$ being the gelation concentration. All dispersions formed stable sols during the initial time, and with aging network-like structures were found to form via two routes: one for $c < c_g$, by *phase separation* and another for $c > c_g$, through *equilibrium gelation*. This has invoked and called for a revisit of the phase diagram of aging MMT dispersions.

Emergence of anisotropic interaction between patchy colloids give rise to an array of unusual and novel soft materials like empty liquids with vanishing density, arrested networks and equilibrium gels which do not necessarily follow an underlying phase separation process to be borne. Recent research in patchy colloids has thrown up possibilities for intelligent design and development of structured nano-assemblies following a bottom-up approach. These systems are inherently rich in dynamics, and a hierarchy of time and length scales adequately defines the properties of the soft matter so formed. Phase stability of dispersions of anisotropic particles has been of much scientific debate in the recent times. Many phase diagrams have been conceived and proposed for colloidal particles having different degree of geometrical anisotropy, namely rods, platelets, disks etc in their dispersion states[1-3]. Clays are discotic platelets with varying aspect ratio and surface charge density, and their dispersions exhibit an array of soft matter phases that evolve with waiting time $t_w$ as well as solid concentration $c$. Thus, the $t_w$-$c$ phase diagram is replete with non-trivial and equilibrium soft matter phases continuously evolving with time. From the application point of view, suspensions of clays offer unique properties, including their ability to form arrested states like gels, glasses and liquid crystalline structures under ambient conditions. Remarkably, compared to spherical colloids, clay platelets exhibit gel and glass phases at very low concentrations, which owe their origin to their unique geometrical structure and surface charge heterogeneity.

In this work we demonstrate the spontaneous evolution of various self-assembled phases from a homogeneous aqueous dispersion of Montmorillonite (MMT, platelet diameter 250 nm and thickness 1 nm) nanoclay platelets based on the observations made over a period of 3.5 years. We have established the $t_w$-$c$ phase diagram for this system in salt-free suspensions in normal pH conditions using rheology experiments and have noticed that these suspensions do undergo nontrivial phase evolution and aging dynamics. We have observed distinctive phase separation, equilibrium fluid and equilibrium gels in the $t_w$-$c$ phase space. Cole-Cole plots derived from rheology measurements indicated the presence of inter connected network-like structures for $c > c_g$, $c_g$ being the gelation concentration. Flow curves exhibited non-Newtonian behavior above $c_g$, therefore gave rise to yield stress $\tau_0$ which grew with concentration of clay, given by $\tau_0 \sim c^\alpha$, $\alpha = 3.0 \pm 0.3$. This observation showed that the yield stress dependence on the concentration of clay was universal and invariant of colloidal aspect ratio. The dispersions with $c > c_g$ were not affected by dilution, which ensured the dominance of electrostatic attraction. During the initial period, all dispersions formed stable sols, and with aging network-like structures were found to form via two routes: one for $c < c_g$, by phase separation and another for $c > c_g$, through equilibrium gelation. This has invoked and necessitated a revisit of the phase diagram of aging MMT dispersions.

Montmorillonite is one of the natural clays, which has high aspect ratio ($\approx 250$) and is a macroscopically swelling, 'active' clay that has the capability for taking up large amounts of water to form stable gels. The phase diagram of MMT is least explored in the literature, though some attempts in that direction have been made in the past[4,5]. But a systematic study of the aging dynamics has not been performed as yet. Recently, we have reported the sol-gel transition behavior in laponite-MMT mixed clay dispersions[6]. On the other hand, visco-elastic properties of many clay suspensions have been extensively probed[7-10]. For instance, rheology behaviour of bentonite slurries as a function of molar ratio of $Na^+/Ca^{+2}$, pH and with an array of additives such as pyrophosphate, polyphosphate[11] and SDS surfactant[12] has been reported. According to Lagaly *et al.* bentonite clay particles through face (−)/edge (+) (FE) attraction form "house of cards" structures in acidic medium and band-like structures are formed through cation-mediated face (−)/face (−) (FF) attraction in alkaline medium[13]. The nature and mechanical strength of these microstructures formed are shown to be influenced by the structure and nature of the adsorbed additives[14]. The EF configuration would be preferred at low pH, i.e where the edges are purportedly having a positive charge, whereas higher pH and higher concentration would prefer the EE configuration. Eventually, high charge density on the edges would tend to prefer the formation of FF-like structures. Recently, a detailed study on the tactoid formation in Ca- Montmorillonite was carried out by Segad et al.[15]. Montmorillonite, natural swelling clay that absorbs water and swells substantially builds a yield stress gel at very low solid concentration, below 4 % (w/v) as compared to Laponite, a synthetic discotic clay with a diameter of ~ 25 and 1 nm thickness, which displays similar behavior at even lower solid concentration of 1 % (w/v)[16,17]. The presence of adsorbed water covering the clay particles produces characteristic cohesive plastic behavior of clay minerals. Recent studies described clay as patchy colloids with limited valance which have the ability to form empty liquids with low coordination number, and equilibrium gels which do not necessarily follow an underlying phase separation process[16,18]. However, no $t_w$-$c$ phase diagram for MMT dispersions has been proposed so far. In this letter, we offer to close the aforesaid knowledge gap by proposing the required phase diagram for the first time.

The flow behavior of any system is characterized from the relationship between the shear stress and the shear rate. In fluid mechanics, the shear rate is defined as the change of shear strain per unit time, and the shear stress is proportional to shear rate with a viscosity function. Therefore the ratio of shear stress to shear rate is called apparent viscosity, which is a measure of resistance to the flow of the fluid under consideration. This is represented as $\eta_a = \tau/\dot{\gamma}$ where $\tau$ is the shear stress and $\dot{\gamma}$ represents the shear rate. The investigated clay dispersions exhibited an apparent viscosity $\eta_a$ that decreased with increase in shear rate up to 1000 s$^{-1}$ due to the rupture of the self-assembled microstructures present in the dispersion. The higher viscosity of dispersion was a result of stronger electrostatic interactions prevailing between the platelets. The flow curves for the examined dispersions are shown in Fig. 1a for $c > c_g$, the data shows pseudoplasticity (shear thinning) having a yield stress defined as the stress above which the material flows like a viscous fluid. This observation indicated that these colloidal dispersions were non-Newtonian materials. The viscosity of the shear thinning fluid could be described by power-law function given by $\eta_a \sim \dot{\gamma}^{-m}$. The exponent, $m$ increased from 0.83 to 1 in the samples with increase in concentration (inset of Fig. 1a). This observation made us look at the growth of yield stress of phase arrested samples deeply.

The noticed stress response of the samples with aging indicated about the emergence of a well defined yield stress. In order to quantify this, measured flow curves were analyzed within the Herschel-Bulkley formalism given by[19], $\tau(\dot{\gamma}) = \tau_0 + K\dot{\gamma}^n$ where $\tau_0$ is the yield stress, $K$ represents the consistency factor and $n$ is a parameter that characterizes the pseudoplasticity of the system. Generally, the yield stress $\tau_0$ is determined as the intercept of the flow curve at zero shear rate (Fig. 1b). The yield stress, $\tau_0$ obtained from the flow curves showed a power-law dependence on clay concentration which is illustrated in Fig. 1b and this data could be correlated to concentration through $\tau_0 \sim c^\alpha$, $c > c_g$; $\alpha = 3.0 \pm 0.3$. The observation of the power-law exponent 3.0 for the gel samples is consistent with the value reported by Pignon et al.[20], and Pujala et al. for Laponite dispersions[17]. This concludes that the yield stress dependence on concentration of the clay is universal and invariant of aspect ratio of the clay particle.

The phase homogeneity in polymer solutions and melts is often derived from Cole-Cole plot where the imaginary part of the complex viscosity $\eta''$ is plotted as function of the real part $\eta'$. Generally, in a melt, at very low frequencies, viscous behaviour is observed whereas at higher frequencies elastic properties prevail. In this formalism, $\eta^*(\omega) = \eta'(\omega) + i\eta''(\omega)$ and the low and high frequency viscosity values are given by $\eta_0$ and $\eta_\infty$ respectively. The Cole-Cole empirical expression is written as[22,23] $\eta^* - \eta_\infty = \dfrac{(\eta_0 - \eta_\infty)}{[1 + (j\omega\tau_{cc})^{1-\alpha}]}$; $0 < \alpha < 1$.

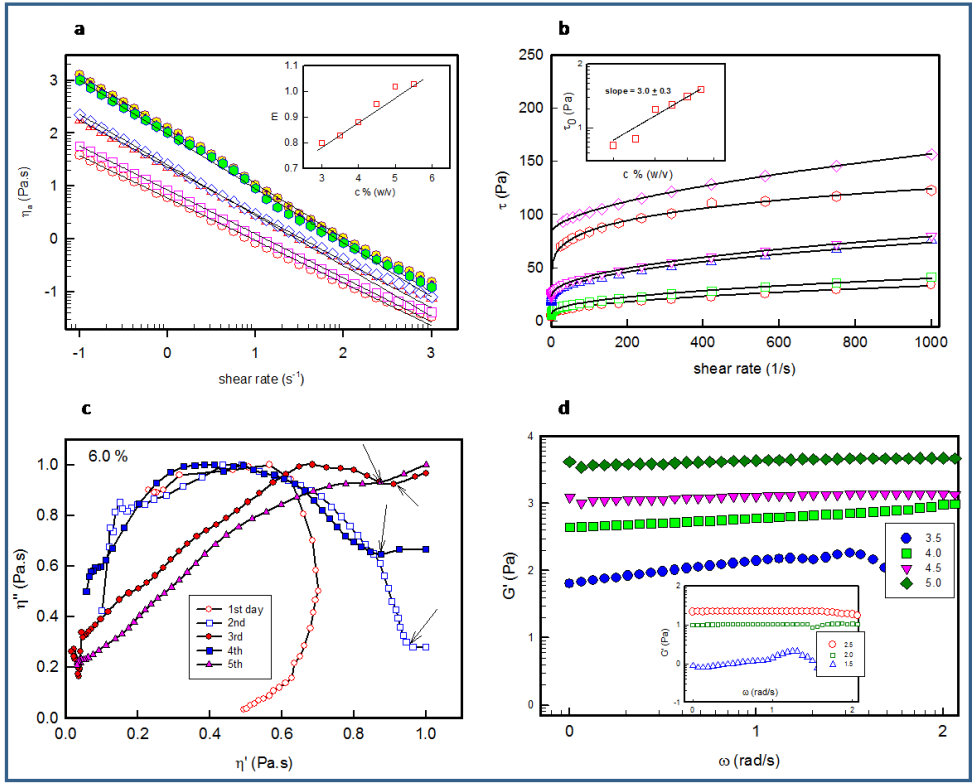

**FIG. 1:** Visco-elastic measurements of aging Montmorillonite dispersions. **(a)** Steady-state viscosity of the system plotted as a function of the shear rate for different clay concentrations (○) 3.0%; (□) 3.5%; (Δ) 4.0%; (◊) 4.5.0%; (⓪) 5.0%; (⊕) 5.5%. The data was least- squares fitted to equation $\eta_a \sim \dot{\gamma}^{-m}$. Inset shows the variation of exponent $m$ with clay concentration. **(b)** Flow curves for the clay samples of different concentrations 3.0, 3.5, 4.0, 4.5 and 5 % (w/v) from bottom to top respectively. The solid lines show fitting to the Herschel-Bulkley model. The value of exponent $n$ is $0.45 \pm 0.02$ and is same for all the curves. Inset shows the variation of yield stress as a function of clay concentration. The power-law exponent determined was $\alpha = 3.0 \pm 0.3$. **(c)** Cole-Cole plots for the specified concentration of MMT with aging. The data were normalized by dividing $\eta'$ with $\eta'_{max}$ and $\eta''$ with $\eta''_{max}$. The arrows indicate the point of upturn. **(d)** Frequency dependent storage modulii of the MMT dispersions having different concentrations after 2 years of sample preparation. Inset shows the same for samples with $c < c_g$, which are performed on the samples drawn from the lower part of the cell.

The aforementioned expression is interpreted as originating from a superposition of several Debye relaxations[22,23]. The mean relaxation time is given by $\tau_{cc}$. This representation has been extensively used to characterize homogeneity of soft matter systems and their composites. For a homogeneous phase, the Cole-Cole plot is a perfect semicircle ($\alpha = 0$) with a well defined relaxation time. Any deviation from this well defined shape indicates mixed dispersion and phase segregation due to immiscibility. Such phases possess a relaxation time spectrum as mentioned earlier. Previous works showed that plotting the two components of dynamic viscoelastic characteristics (modulus or viscosity) against each other yields an arc shaped curve if the process can be depicted with a single relaxation time[21-26]. The arc transforms to a semicircle or a skewed semicircle if the material owns a distribution of relaxation times[23,24]. The presence of more than one process with different relaxation times needs further modification in the so called Cole-Cole plot, i.e., having a new semicircle or the appearance of a tail [25,26]. The Cole-Cole plots of our dispersions are shown in Fig. 1c for the aging samples. We have employed the rheology to monitor network rigidity of the arrested phase. It was observed that the elastic modulus is higher than the viscous modulus for $c > c_g$ as shown in Fig. 1d. The samples drawn from the lower part of 2.0 and 2.5 % (w/v) dispersions showed viscoelastic behavior whereas the same taken from 1.0 and 1.5% (w/v) samples showed viscous property.

Initially, the dispersions were homogeneous which is evident from the nearly semicircular profile of the plot. As the sample aged, an unambiguous tail started appearing which indicated network formation in these systems[21-26]. In a diverse example, the silicate network was detected by the increase in the complex viscosity and storage modulus with decreasing frequency, in the small frequency range of the mechanical spectrum[22-25]. The increase in the elastic component of the viscosity and modulus is most probably due to the formation of a network structure and the network obviously deforms with different relaxation times than the homogeneous melt, thus we expect a deviation from a semicircle in the afore-mentioned representation. The deviation from a skewed semicircle is clearly visible as the sample aged. The change in the shape of the plots indicated the appearance of a new relaxation processes, likely the formation of the arrested networks (Fig. 1c).

As the sample aged the size of the aggregates started growing due to EF or FF interactions at lower concentrations. The particles or the aggregates explore the entire phase space and interact and form aggregates of larger size that eventually phase separate due to gravity. The phenomenon that we have observed is quite similar to Laponite system reported recently by Ruzicka and co-workers[16]. Samples undergo an extremely slow, but clear, phase-separation process into clay-rich and clay-poor phases that are the colloidal analogue of vapor-liquid phase separation. Figure 2a depicts the picture of phase separation at lower concentrations below $c_g$ and the formation of equilibrium gels above $c_g$. Stunningly, the phase separation terminates at a finite but very low clay concentration, above which the samples remain in a homogeneous arrested state, which may be called an empty liquid[16, 18].

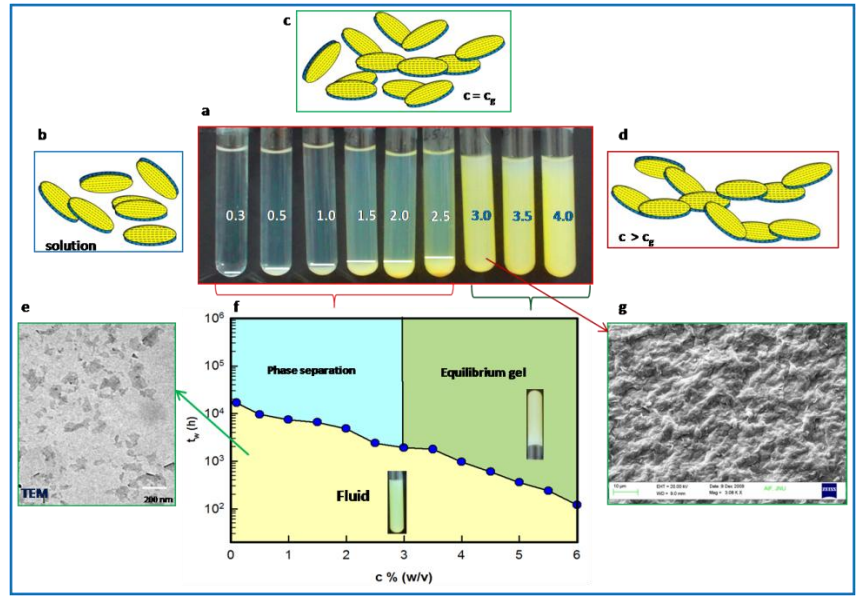

**FIG. 2:** Phase Diagram of aging MMT dispersions. **(a)** Series of MMT dispersions indicating the phase separation states below $c_g$ (= 2.5%) and the equilibrium gels above $c_g$. This photograph was taken after 3.5 years. Note that the height of the equilibrium line increases with the clay concentration below $c_g$. Clay concentrations are in % (w/v). **(b-d)** Cartoon of self-assembly clay microstructures. **(e)** TEM micrograph of MMT in dilute region. **(f)** Phase diagram of MMT with concentration of clay versus the aging time or the phase separation time. Three regions were clearly identified: stable fluid, phase separation and equilibrium gel. **(g)** SEM image of MMT dispersion (3% (w/v)).

Now the question the arises: what is the kind of interaction present that gives rise to the formation of network-like self-assembled structures under the present experimental conditions? Goh et al. have reported that for bentonite slurries there are three linear regions in yield stress versus zeta potential data [27]. Two linear regions with negative slope were observed where a steepest slope was found in low pH and a moderate slope was seen in the intermediate pH region. A positive slope was noticed for pH > 7. The positive slope region was ascribed to anisotropic charge attraction between the negative face and positive edge[27]. Obviously, for this to occur, the nature of the platelet edge charge must still be positive in high pH dispersions. The most likely self-assembled microstructure is therefore conceived to be the "overlapping coins" or the band-like structure (Figs. 2b-d). Recently Cui et al. found the existence of aggregates at pH ≈ 9 [15]. The aggregates however form, presumptively due to the distribution of charges in and around the clay particles, leading to quadrupole (and higher order multi-pole) interactions between particles. They have found existence of the strand-like structures in MMT suspensions.

In another study, Jonsson et al. modeled the free energy of interaction between two nanometric clay platelets plunged in an electrolyte solution each plate contains negative charges on the face, and the edge, positive charges in face–edge (FE), overlapping coin and staked platelet configurations using a Monte Carlo simulation[28]. Their calculations predicted that the "overlapping coin" configuration gave rise a global free energy minimum at intermediate salt concentrations. In a compression study by Callaghan et al. on sodium bentonite (montmorillonite) slurry, the clay particles initially formed a disordered gel but the first compressions leads the formation of "overlapping coin" or parallel plate array structure regardless of the NaCl concentrations[29]. TEM micrograph indicates the presence of individual clay particles in the dilute dispersions shown in Fig. 2e. But with waiting time and at $c > c_g$ the formation of network structures is visible from the SEM image (Fig. 2g).

We have observed three clear phase states in the the dispersions: (i) stable fluid, (ii) phase separated dispersion and (iii) equilibrium gel. Thus, we propose a phase diagram for these dispersions with concentration of clay versus waiting time as variables which is shown in Fig. 2f.

We have undertaken a comprehensive study to map the phase diagram of MMT dispersions extended over a time span of 3.5 years. The system exhibited three distinct phase states: (i) initially stable homogeneous solution, (ii) with aging the particles self-assembled and phase separated for concentrations $c < c_g$ and (iii) in the third phase the dispersions entered into an equilibrium gel phase for $c < c_g$, where the self-assembled 3-D networks formed through overlapping coin packing interactions. These samples exhibited aging behavior which were captured through rheology measurements. Most recently similar observations were made for aging Laponite dispersions, which exhibited presence of empty liquids, equilibrium gels and glasses during the course of waiting. Interestingly, these did not necessarily follow an underlying phase separation process. It may be concluded that the phenomenon of existence of phase separated solution and formation of equilibrium gels is ubiquitous in anisotropic colloidal particles associated with inhomogeneous charge distribution. Remarkably the aforesaid phenomena was independent of platelet aspect ratio. Most recently Meneses-Juarez and co-workers have observed the formation of a liquid phase with vanishing density, called an empty liquid [18]. Based on the data in hand, we are unable to vouch for the existence of arrested empty liquids in our $t_w$-$c$ phase diagram. Since such arrested empty liquid states are formed under reduced valence and low coordination number conditions in aging dispersions spontaneously due to the patchy distribution of charge on the platelet surface, their signatures could be better captured in structure factor measurements. As of now, empty liquid phase has been experimentally observed in laponite dispersions only. Thus, the comprehensive understanding of the kinetics of dynamic arrest and phase separation in the discotic colloidal suspensions continues to remain a challenging and poorly explored problem.

**Materials and Methods**
**MMT Sample Preparation**

Na MMT (*Na* Cloisite), a hydrated aluminium silicate, was purchased from Southern Clay Products, U.S. and used as received. Chemically it is hydrated sodium calcium aluminium magnesium silicate hydroxide (*Na*, *Ca*)$_{0.33}$(*Al*, *Mg*)$_2$(*Si*$_4$*O*$_{10}$)(*OH*)$_2$·$nH_2O$. The powder was first dried in an oven at 300$^0$C to remove the moisture content. An aqueous dispersion of MMT was prepared by dispersing it in deionized water at pH 8.5. The sample was stirred vigorously using magnetic stirrer for 48 hours to ensure complete dissolution of the clay particles and the fractionation was done as described in ref. [30]. The samples were prepared in a clean environment and exposure to air was avoided by sealing them in tight vials. We have carried out experiments in the broad concentration range 0.1 - 6 % (w/v). All the experiments were performed under room temperature conditions, temperature 25$^0$ C and relative humidity < 40%. For reproducibility of the results we have prepared three sets of samples with a gap of one month for each set and performed the measurements.

**Rheology** experiments were performed using an AR-500 model stress controlled rheometer (T.A. Instruments, UK). Frequency sweep experiments were performed using cone-plate geometry (4 cm diameter, 2$^0$ cone angle and 50 mm truncation gap) with the oscillatory stress value set at 0.1 Pa specially to observe low frequency behavior of storage and loss modulii. The time dependent viscosity of the dispersions was measured by Sine-wave vibro viscometer (SV-10 model; A & D Company, Japan). The instrument measures viscosity by detecting the driving electric current necessary to resonate the two sensor plates at constant frequency of 30 Hz and amplitude of less than 1mm, which ensures low shear measurement conditions.